\documentclass[aps,prb,twocolumn,floatfix]{revtex4-1}
\usepackage{times} 
\usepackage{epsfig} 
\usepackage{amsmath}
\usepackage{amssymb} 
\usepackage{graphicx}
\usepackage{dsfont}
\usepackage[colorlinks=true,citecolor=red,linkcolor=blue]{hyperref}

\newcommand{\bfk}{\mathbf{k}}

\begin{document}

\title{Dynamical recovery of SU(2) symmetry in the mass-quenched Hubbard model}
\author{Liang Du}
\affiliation{Department of Physics, The University of Texas at Austin, Austin, TX 78712, USA}
\author{Gregory A. Fiete}
\affiliation{Department of Physics, The University of Texas at Austin, Austin, TX 78712, USA}
\begin{abstract}
We use non-equilibrium dynamical mean-field theory with iterative perturbation theory as an impurity solver to study the recovery of $SU(2)$ symmetry in real-time following a hopping integral parameter quench from a mass-imbalanced to a mass-balanced single-band Hubbard model at half-filling.  A dynamical order parameter $\gamma(t)$ is defined to characterize the evolution of the system towards $SU(2)$ symmetry. By comparing the momentum dependent occupation from an equilibrium calculation (with the $SU(2)$ symmetric Hamiltonian after the quench at an effective temperature) with the data from our non-equilibrium calculation, we conclude that the $SU(2)$ symmetry recovered state is a thermalized state. Further evidence from the evolution of the density of states supports this conclusion. At the same time, we find the order parameter in the weak Coulomb interaction regime undergoes an approximate exponential decay. We numerically investigate the interplay of the relevant parameters (initial temperature, Coulomb interaction strength, initial mass-imbalance ratio) and their combined effect on the thermalization behavior. Finally, we study evolution of the order parameter as the hopping parameter is changed with either a linear ramp or a pulse.  Our results can be useful in strategies to engineer the relaxation behavior of interacting, quantum many-particle systems.
\end{abstract}
\date{\today}
\maketitle
\section{Introduction}
Research on non-equilibrium quantum phase transitions have seen dramatic progress and in the past decade.\cite{Lindner:nat11,Gull:rmp11,Wang:sci13,Aoki:rmp14} Many experimental and theoretical studies have focused on pump-probe experiments in solid state systems where the system is driven out-of-equilibrium by a pump laser,\cite{Manmana:prl07,Rigol:prl07,Kollath:prb07} or in cold atom systems driven\cite{Mentink:prl14,Mentink:nc15,Eckstein:scirep16,Bukov:prl16} by Coulomb interaction strength change. In a real-time quantum phase transition, symmetry breaking or recovery often plays an important role.  For example, the phase transition between paramagnetic and anti-ferromagnetic\cite{Werner:prb12,Tsuji:prl13,Tsuji:prb13} behavior as a function of Coulomb interaction quench is associated with the spontaneous symmetry breaking of lattice symmetries and time-reversal; the transition to a Floquet topological insulator in Bi$_2$Se$_3$,\cite{Lindner:nat11,Wang:sci13} Graphene,\cite{Oka:prb09} and bilayer (LaNiO$_3$)$_2$/(LaAlO$_3$)$_N$ thin films\cite{Du:prb17a,Du:prb17b} is triggered by the time-reversal symmetry breaking induced by shinning with a circularly polarized laser but does not entail any corresponding breaking of lattice symmetries.

If the electronic system driven out-of-equilibrium is strongly correlated, theoretical and numerical techniques to deal with the system are limited, which makes this problem especially challenging.  The Hubbard model is widely considered the simplest model to capture the most essential features of strongly correlated systems (either solid state materials or cold atom systems). Non-equilibrium dynamical mean-field theory (DMFT),\cite{Georges:rmp96,Gull:rmp11,Aoki:rmp14} in which the original lattice problem is mapped onto a Anderson impurity problem with a self-consistently determined bath, has proved to be a powerful tool in solving the Hubbard model.  The non-equilibrium real-time evolution of physical observables can be obtained within the framework of DMFT. 

Even though the methods to solve the Anderson impurity model in equilibrium are well developed, solving the model out-of-equilibrium remains difficult and is still under development.\cite{Wolf:prb14,Balzer:prb15,Cohen:prl15,Dong:prb17} 
Numerically, the robust impurity solver in equilibrium, hybridization expansion continuous time quantum Monte Carlo, can suffer from the dynamical sign problem and the simulation time is usually rather short. The numerically exact impurity solver, weak coupling continuous time auxiliary-field quantum Monte Carlo, is limited to a single-band model at half-filling and a short time evolution. The dynamical sign problem increase exponentially with evolution time, which limit its applicability. Analytically, the iterative perturbation theory (IPT) impurity solver at weak and the non(one)-crossing approximation (NCA, OCA) at strong Coulomb interactions have been shown to be powerful impurity solvers to capture the physical picture in the weak and strong Coulomb interaction regimes, respectively.\cite{Tsuji:prb13,Eckstein:prb10b}

Due to the limitations of the impurity solvers for non-equilibrium DMFT, the Falicov-Kimball (FK) model provides a first attack on strongly correlated electronic systems because the projected impurity model is exactly solvable and provides important information on the Hubbard model.\cite{Freericks:prl06,Eckstein:prl08}.
The FK model is well studied both in  equilibrium and out-of-equilibrium.\cite{Freericks:rmp03,Freericks:prl06} This raises the questions, ``What is the physical connection between the FK model and the Hubbard model?" and ``To what extent does the FK model reveal the physics of the Hubbard model?". To answer these questions, previous works focused on studying the two models separately, and compared the results of each model with each other.  

For cold atom systems, Eckstein {\it et al.}\cite{Eckstein:prl08,Eckstein:prl09,Eckstein:prb10a} studied the evolution of physical quantities as a function of time driven by a Coulomb interaction quench in the FK model and the Hubbard model, respectively.  For the FK model driven by a Coulomb interaction quench, a non-thermal steady state (not a thermalized state) results that can be statistically described by the generalized Gibbs ensemble.\cite{Eckstein:prl08} For the Coulomb interaction quenched Hubbard model, a pre-thermalization behavior and dynamical phase transition are observed.\cite{Eckstein:prl09} For a solid state system driven by a constant electric field, Freericks {\it et al.}\cite{Freericks:prl06,Freericks:prb08} studied the FK model and found damped Bloch oscillations. Eckstein {\it et al.}\cite{Eckstein:prl11} studied the Hubbard model: except for the damped Bloch oscillation observed in FK model, the current decays to zero and remains there. Fotso {\it et al.}\cite{Fotso:scirep14} compared the thermalization behavior of the FK model and the Hubbard model driven by a DC electric field. The FK model can have one of two generic evolution behaviors: (1) either monotonic or oscillatory approach to an infinite-temperature steady state or (2) either monotonic or oscillatory approach to a non-thermal steady state.
In addition to the above two features, the Hubbard model can exhibit an extra feature by evolving to an oscillatory state. 

In contrast to previous studies, in this work we would like to build the connection between the FK model and the Hubbard model by quenching the hopping parameter of the frozen species (the one that is not able to hop on the lattice) in the FK model to the Hubbard model. In order to avoid the singularity of the FK model (bandwidth for one species is zero), we use the mass-imbalanced Hubbard model with large hopping asymmetry and study the time-dependent evolution of observables following a quench between the mass-imbalanced and mass-balanced Hubbard model.  Here mass-imbalance (mass-balance) means the spin-$\uparrow$ and spin-$\downarrow$ hopping parameter are unequal (equal) to each other.  Previous work has studied in the hopping parameter quench of the Hubbard model for equal strength hopping of two spin species, and can be solved as an Coulomb interaction quench problem with scaled time.\cite{Tsuji:prl11}  In this work, we show a quench on only one hopping parameter leads to rather different physics.

One of the central results of our work is that a dynamical phase transition appears.  To put our results in context, it useful to summarize related work that also found dynamical phase transitions.  By quenching the Coulomb interaction between two different phase regimes in equilibrium, Tsuji {\it et al.}\cite{Tsuji:prl13} studied the dynamical phase transition between an antiferromagnetic and paramagnetic state. Two dynamical transition points are observed with one the thermal transition and the other related to a non-thermal antiferromagnetic phase. 

By contrast, we study the evolution of $SU(2)$ symmetry recovery by quenching from the mass-imbalanced to mass-balanced Hubbard model. This can be experimentally realized in cold atom systems by tuning the lattice potential amplitude and the recoil energy.\cite{Nguyen:prb15}  In the mass imbalanced Hubbard model, the $SU(2)$ symmetry is broken. By quenching the hopping integral of one spin species to be the same as the other one, the Hamiltonian recovers its $SU(2)$ symmetry. However, the evolution of physical observables as a function of time remains unclear. Our work fills that gap. We address the following questions: (1) Is the $SU(2)$ symmetry recovered state the same as the equilibrium thermalized state? (2) What is the dependence of the evolution process on the Coulomb interaction, temperature, and the initial mass imbalance? (3) How does the time-evolution change if we set up the quench process as a linear ramp or pulse shape? 

In this work, we show there an $SU(2)$ order parameter can indeed serve as a criteria for a dynamical phase transition. As the $SU(2)$ symmetry in observables is recovered following the quench, the system is thermalized at the same time. We show that the evolution of the $SU(2)$ order parameter has a monotonic dependence on the mass imbalance, temperature, and Coulomb interaction. The pulse shape influences the time evolution.

Our paper is organized as follows. In Sec.\ref{sec:model}, we describe the mass-imbalanced Hubbard model and illustrate how we calculate several physical observables within time-dependent dynamical mean-field theory, such as the momentum dependent occupation.  We also define the order parameter we use to characterize the $SU(2)$ symmetry. In Sec.\ref{sec:benchmark}, we characterize the $SU(2)$ symmetry recovered state as a thermalized state. The dependence on the Coulomb interaction, the initial temperature, and the initial mass-imbalance ratio is studied. We compute the evolution of the order parameter for a linear ramp and a pulse change of the hopping parameter in Sec.\ref{sec:pulse}. Finally, in Sec.\ref{sec:conclusion} we summarize the main conclusions of this work.

\section{Model and Method}
\label{sec:model}
The time-dependent mass-imbalanced single-band Hubbard model at half-filling is given by,\cite{Dao:pra12,Liu:prb15,Philipp:epjb17,Sekania:prb17,Du:prb17c}
\begin{align}
   H(t) = &\sum_{\langle ij \rangle\sigma} -V_\uparrow c^\dagger_{i\uparrow} c_{j\uparrow}^{} - V_\downarrow(t) c^\dagger_{i\downarrow} c_{j\downarrow}^{} + h.c. \nonumber\\
       &+ U \sum_i \left(\hat n_{i\uparrow}^{} -\frac{1}{2}\right)\left(\hat n_{i\downarrow}^{}-\frac{1}{2}\right),
\label{eq:ham}
\end{align}
where $c^\dagger_{i\sigma}$ ($c_{i\sigma}$) creates (annihilates) an electron at site $i$ with spin $\sigma$, and $\hat n_{i\sigma} = c^\dagger_{i\sigma} c_{i\sigma}$ is the corresponding density operator. The notation $\langle ij\rangle$ indicates the hopping is restricted to nearest neighbors, $V_{\downarrow}(t)$ ($V_{\uparrow}$) is the time dependent (independent) hopping integral parameter for spin-$\downarrow$ (spin-$\uparrow$) electron ($t$ is reserved to denote time).  The time dependence of the hopping parameter can be introduced in optical lattices through lasers. Here $U$ denotes the time independent Coulomb interaction strength between spin-$\uparrow$ and spin-$\downarrow$ fermions occupying the same site.
Throughout this paper, we fix the spin-$\uparrow$ hopping integral to be time independent $V_{\uparrow} = 1$ and set $V_{\uparrow}$ ($1/V_{\uparrow}$) as our unit of energy (time). The mass imbalance $r = V_\downarrow(t)/V_\uparrow$ is restricted to lie between 0 and 1. The system is initially prepared in the thermal equilibrium state of the mass imbalanced Hubbard model with $V_{\downarrow}(t<0) \neq V_{\uparrow}$ and finite repulsive Coulomb interaction $U > 0$. Here the $SU(2)$ symmetry of the system is broken. The quench dynamics are studied by fixing the Coulomb interaction $U$ to be finite while quenching the spin-$\downarrow$ hopping integral to be $V_{\downarrow}(t \geq t_q) = V_{\uparrow}$ from an initial $V_{\downarrow}(t<0) \neq V_{\uparrow}$ state, where $t_q$ is the ramp time of the hopping parameter change.

We consider a Bethe lattice, which has a semi-elliptic density of states, 
\begin{equation}
\label{eq:dos}
      \rho_\sigma(\epsilon) = \frac{1}{2\pi V_\sigma^2} \sqrt{4V_\sigma^2 - \epsilon^2},
\end{equation}
with half bandwidth $D_\sigma = 2 V_\sigma$.
The mass-imbalanced Hubbard model (\ref{eq:ham}) can be solved exactly using non-equilibrium dynamical mean field theory (DMFT),\cite{Georges:rmp96,Freericks:prl06,Eckstein:prl09,Gull:rmp11,Aoki:rmp14} which maps the lattice model self-consistently onto a single-site Anderson impurity model. We use non-equilibrium dynamical mean field theory with iterative perturbation theory as an impurity solver to solve the mass imbalanced Hubbard model at finite temperature.
We enforce the paramagnetic solution and half-filling of both spin-$\uparrow$ and spin-$\downarrow$ electrons. In the Hubbard model, these constraints can be fulfilled by explicitly symmetrizing over the two spin spices and setting the chemical potential to be $\mu = U/2$, respectively. Away from this mass balanced Hubbard model limit, we again enforce half-filling by fixing $\mu = U/2$. However to ensure the paramagnetic solution at half-filling, we symmetrize the Weiss's functions in the Keldysh time contour using particle-hole symmetry: 
$\mathcal{G}_{0,\sigma}(t, t') = -\mathcal{G}_{0,\sigma}(t',t)$.
The DMFT self-consistent condition for the Bethe lattice\cite{Eckstein:njp10} density of state is
\begin{equation}
     \Delta_{\sigma}(t,t') = V_{\sigma}(t) G_{\sigma}(t,t')V_{\sigma}(t').
\end{equation}
\begin{figure}[t]
\includegraphics[width=1.0\linewidth]{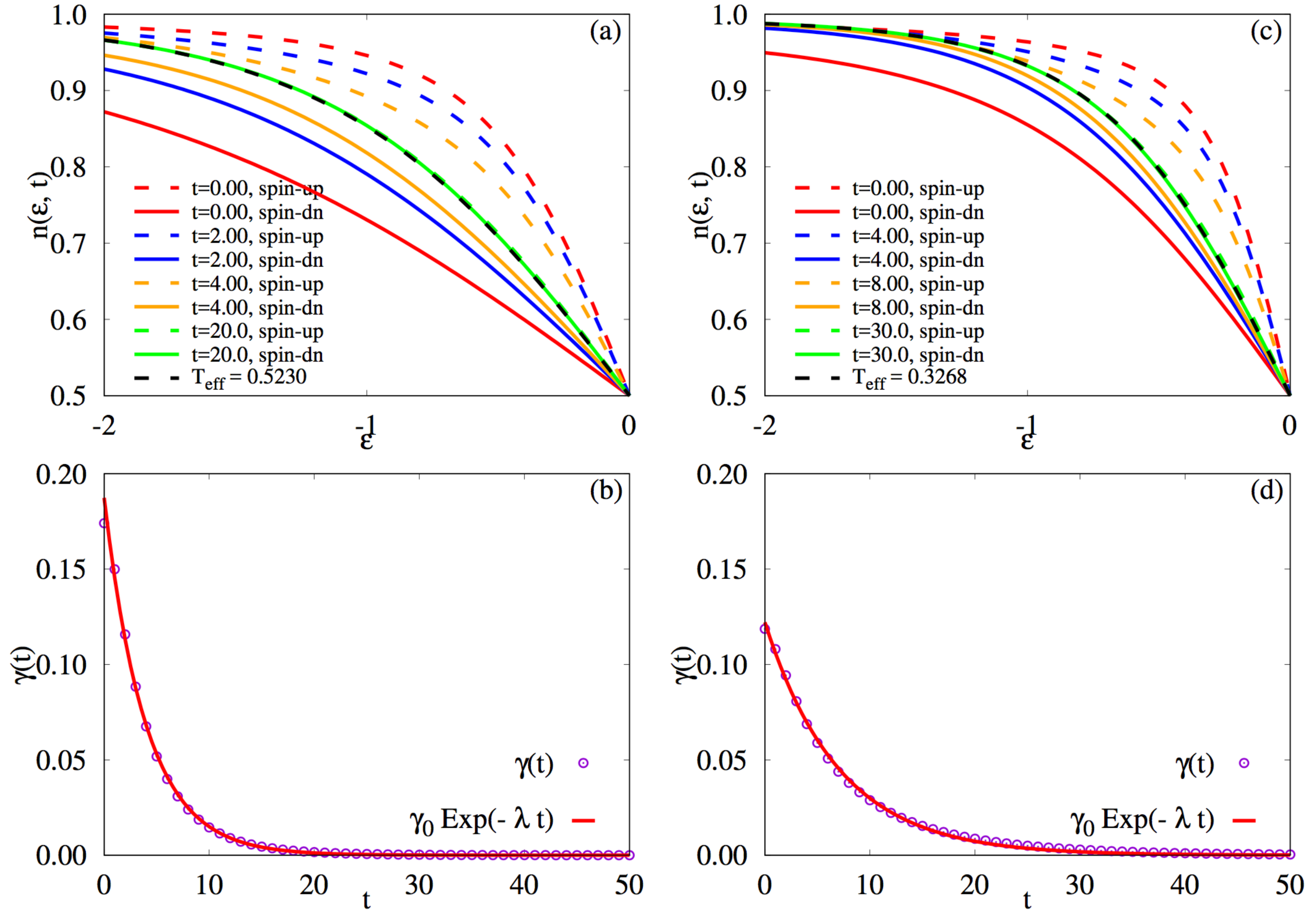}  %
\caption{(Color online) The spin-resolved momentum dependent occupation as a function of energy at different times for the quenched mass-balanced Hubbard model at half-filling. The mass-imbalance at time $t=0^-$ is fixed at $r_0 = V_\downarrow/V_\uparrow=1/4$. The Coulomb interaction is fixed at $U=1.0$. The fitting function used for order parameter is $\gamma(t) = \gamma_0 e^{-\lambda t}$.
(a) spin-resolved momentum distribution at $\beta=5.0$,
(b) the order parameter as a function of time at $\beta=5.0$, the fitting parameters are $\gamma_0=0.1874, \lambda=0.2527$.
(c) spin-resolved momentum distribution at $\beta=10.0$,
(d) the order parameter as a function of time at $\beta=10.0$, the fitting parameters are $\gamma_0=0.1221, \lambda=0.1408$.
We used solid and dashed lines to stand for $n(\epsilon, t)$ with spin-$\uparrow$($\downarrow$) electrons.
The black dashed line is the equilibrium calculation with the half-filled quenched Hamiltonian at effective temperature $T_{\mathrm{eff}}=0.5230 (\beta_{\mathrm{eff}}=1.912)$ (a) and $T_{\mathrm{eff}} = 0.3268 (\beta_{\mathrm{eff}}=3.060)$ (c), where the momentum dependent spin-$\uparrow$ and spin-$\downarrow$ occupation is the same.
}
\label{fig:nkt}
\end{figure}
\begin{figure}[t]
\includegraphics[width=1.0\linewidth]{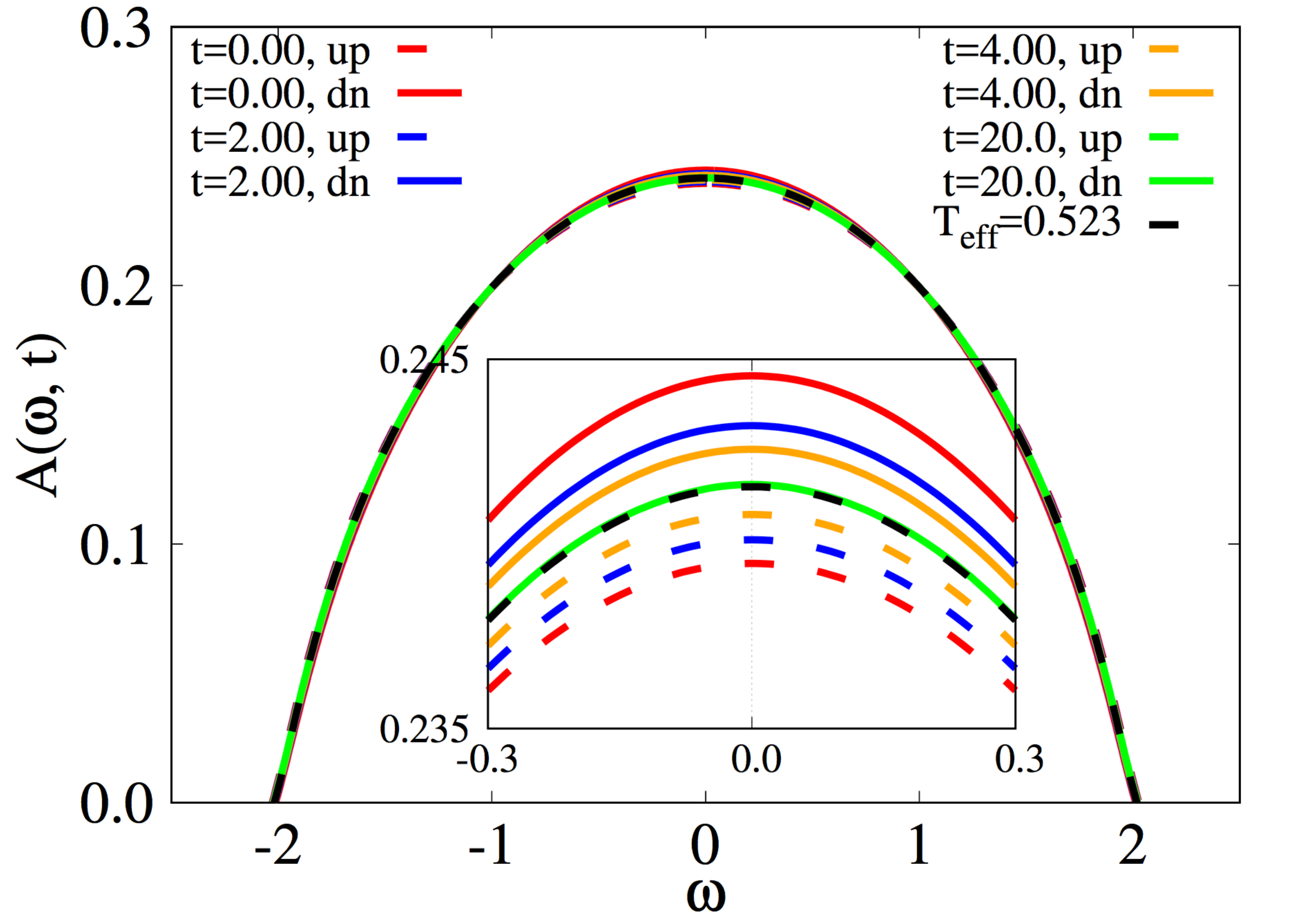}  %
\caption{(Color online) The spin-resolved density of states $A_\sigma(\omega, t)$ as a function of energy at different times for the quenched mass-balanced Hubbard model at half-filling. The mass-imbalance at time $t=0^-$ is fixed at $r_0 = V_\downarrow/V_\uparrow=1/4$. The inverse temperature and Coulomb interaction are fixed at $\beta=5.0$ and $U=1.0$. 
The spin-resolved densities of states are plotted at $t=0,2,4,20$.
We used solid and dashed lines to stand for $A_\sigma(\epsilon, t)$ with spin-$\uparrow$($\downarrow$) electrons.
The black dashed line is the equilibrium calculation with the half-filled quenched Hamiltonian at effective temperature $T_{\mathrm{eff}}=0.5230 (\beta_{\mathrm{eff}}=1.9120)$ where the density of states with spin-$\uparrow$ and spin-$\downarrow$ occupation are the same. The inset shows the zoomed part with $\omega \in [-0.3, 0.3]$.
}
\label{fig:dost}
\end{figure}
\begin{figure*}[t]
\includegraphics[width=1.0\linewidth]{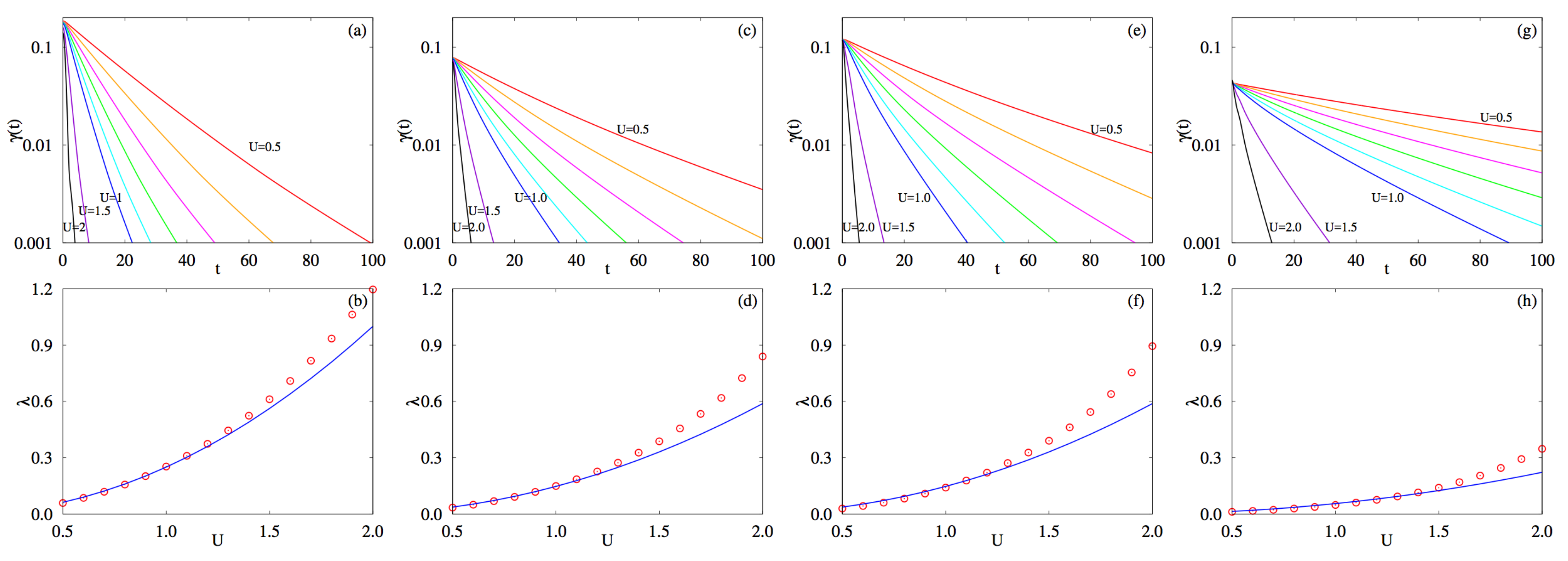}  %
\caption{(Color online) Order parameter $\gamma(t)$ as a function of time for the half-filled mass imbalanced Hubbard model($r_0=V_\downarrow/V_\uparrow=1/4,1/2$) after a hopping integral quench from $r_0=1/4,1/2$ to $r=1$ at $U=0.5,0.6,0.7,0.8,0.9,1.0,1.5,2.0$.
(a-b) $r_0=1/4$ and $\beta=5.0$, (c-d) $r_0=1/2$ and $\beta=5.0$
(e-f) $r_0=1/4$ and $\beta=10.0$, (g-h) $r_0=1/2$ and $\beta=10.0$.
The approximate decay rate as a function of Coulomb interaction strength is plotted as open dots in the bottom panes with the solid line (quadratic function) as a guide to the eye.
}
\label{fig:gqch_u}
\end{figure*}
The expectational value of an observable $\mathcal{O}$ at time $t$ is given by,
\begin{align}
      \langle \mathcal{O}(t) \rangle = \frac{1}{Z_0} \mathrm{Tr}[e^{-\beta H(t<0)}U(0,t)\mathcal{O}U(t,0)],
\end{align}
where $Z_0$ is the partition function of the non-interacting Hamiltonian at $t<0$, $U(t,0) = \mathcal{T}\exp[-i\int_{0}^{t}H(t)dt]$ is the time evolution operator.
The momentum dependent density matrix is written as
\begin{align}
     n_{{\bf k}\sigma}(t) = n_\sigma(\epsilon_{\bf k},t) =  -i G_{{\bf k}\sigma}^{<}(t,t),
\end{align}
where $G_{{\bf k}\sigma}^{<}(t,t)$ is the lesser Green's function at equal time $t$.
The momentum-dependent occupation depends only on $\epsilon_{\bf k}$ because the self-energy is momentum independent.
The kinetic energy is  given by
\begin{equation}
     E_{\mathrm{kin}} = \sum_{\sigma}\int d\epsilon_\sigma \rho_{\sigma}^{}(\epsilon_\sigma^{}) \epsilon_\sigma.
     \label{eq:ekin}
\end{equation}
The Coulomb interaction energy is given by
\begin{align}
E_{\mathrm{int}} &= U \langle n_{i\uparrow}(t)n_{i\downarrow}(t)\rangle \nonumber\\
     &= 
 -i\int_{\mathcal{C}}d\bar{t} \Sigma_{ii\uparrow}(t,\bar{t}) G_{ii\uparrow}(\bar{t},t) + \langle n_{i\uparrow}(t)\rangle/2,
 \label{eq:eint}
\end{align}
where $\mathcal{C}$ denotes the Keldysh contour.\cite{Eckstein:prb10a} 
The total energy is 
\begin{equation}
     E_{\mathrm{tot}} = E_{\mathrm{kin}} + E_{\mathrm{int}}.
     \label{eq:etot}
\end{equation}
The real frequency represented retarded Green's is function is 
\begin{equation}
     G_\sigma^{\mathrm{R}}(\omega, t) = \int_{0}^{\infty} ds e^{i(\omega+i 0^+)s}G_\sigma^{\mathrm{R}}(t+s, t).
     \label{eq:gretw}
\end{equation}
The density of states are calculated from the exact relation
\begin{equation}
     A(\omega, t) = -\frac{1}{\pi} \mathrm{Im}G^{\mathrm{R}}(\omega, t).
\end{equation}
The dynamical order parameter we use to characterize the $SU(2)$ symmetry is,
\begin{align}
     \gamma(t \geq t_q) &= \frac{1}{N_\bfk}\sum_{\bfk}|S_{\bfk}^{z}(t)| \nonumber\\
     &= \frac{1}{N_\bfk}\sum_{\bfk}|n_{\bfk\uparrow}(t) - n_{\bfk\downarrow}(t)|\nonumber\\
     &= \int d\epsilon \rho_\uparrow(\epsilon) |n_{\uparrow}(\epsilon, t) - n_{\downarrow}(\epsilon, t)|,
     \label{eq:order}
\end{align}
where $t_q$ is time of the hopping integral ramp to $V_{\uparrow} = V_{\downarrow}$, and $\rho_{\uparrow}(\epsilon) $ is the semi-elliptic density of states defined in Eq.\eqref{eq:dos} with $V_{\uparrow}=1.0$.
If the quenched system is thermalized after long enough time, the effective temperature is calculated by numerically solving the equation,\cite{Eckstein:prb10a}
\begin{equation}
 E(0^+)=\frac{\mathrm{Tr}\left[H(0^+)e^{-\beta_{\mathrm{eff}}H(0^+)}\right]}{\mathrm{Tr}\left[e^{-\beta_{\mathrm{eff}}H(0^+)}\right]},
\label{eq:tempeff}
\end{equation}
where $E(0^+)$ is the same as Eq.\eqref{eq:etot}, and $H(0^+)$ is the Hamiltonian after quench.

\section{thermalization driven by  hopping quench}
\label{sec:benchmark}
The prethermalization behavior in the paramagnetic case occurs when the momentum integrated quantities (Coulomb interaction and spin-resolved kinetic energy) thermalize faster than the momentum dependent quantities (momentum dependent distribution).\cite{Tsuji:prl13} Thus, we must study the momentum dependent observables as a function of time to determine if the system is thermalized. Depending on the system details, there may be convenient quantities for studying the thermalization.  For example, in the Coulomb interaction quench at zero temperature problem, the Fermi-surface dis-continuity in the momentum-dependent occupations can serve as a good criteria because the Fermi-surface jump disappears at the thermalized finite temperature.\cite{Eckstein:prl09} In our calculation we use the momentum integrated order parameter $\gamma(t)$ in Eq.\eqref{eq:order} for $SU(2)$ symmetry as the specific criterion in our case. 

In Fig.\ref{fig:nkt}, we plot the spin-resolved momentum dependent occupation number as a function of energy at different times with fixed Coulomb interaction $U=1.0$ and initial mass-imbalance ratio $r_0=1/4$. 
Here, for the purpose of better vision, we only plot half of energy axis ($\epsilon<0$). The other part of momentum distribution ($\epsilon >0$) is constrained by $n_\sigma(\epsilon,t)+n_\sigma(-\epsilon, t)=1$ which is hold by particle-hole symmetry.
The only parameter difference is $\beta=5.0$ for Fig.\ref{fig:nkt}(a) and $\beta=10.0$ for Fig.\ref{fig:nkt}(c). The corresponding effective temperature for the final thermalized state is calculated using Eq.\eqref{eq:tempeff}.  We find $T_{\mathrm{eff}} = 0.5230$ for case (a) and $T_{\mathrm{eff}} = 0.3268$ for case (c), respectively. 

At time $t=0$, the momentum distribution for spin-$\uparrow$ and spin-$\downarrow$ electrons are apparently separated. The black dashed line is the thermalized value of the momentum distribution with $SU(2)$ symmetry. The area encapsulated by spin-$\uparrow$ and spin-$\downarrow$ distribution is defined as the order parameters breaking the $SU(2)$ symmetry. As time evolves, the area is diminished monotonically. (See $t=2$ and 4 in Fig.\ref{fig:nkt}, for example.) Finally, at $t=20.0$, the area vanishes which indicates the $SU(2)$ symmetry of the Hubbard model is fully recovered in the time evolution of the states. By comparing the $SU(2)$ recovered distribution with the thermalized state at $T_{\mathrm{eff}} = 0.5230$, one sees they match each other, indicating that the $SU(2)$ symmetry recovered state is just the thermalized state.  Therefore, the order parameter defined in Eq.\eqref{eq:order} can serve as a measure of whether the state is thermalized. In Fig.\ref{fig:nkt} (b), we plot the order parameter $\gamma(t)$ as a function of time. We realize the evolution of the order parameter fits an exponential decay reasonably well: $\gamma(t) = \gamma_0 e^{-\lambda t}$. With $\gamma_0=0.1874$ and $\lambda=0.2507$, the fitting function is a good approximation of the original data. Figs.\ref{fig:nkt}(c-d) give very similar information except some quantitative difference, mainly the small initial order parameter $\gamma(t=0)$ and small decay rate $\lambda$. A systematic discussion of these difference is deferred to future sections of this paper.

To confirm our conclusion that the $SU(2)$ symmetry recovered state is a thermalized state, we plot the spin-resolved density of states at different times in Fig.\ref{fig:dost}. Here the density of states is calculated by Fourier transforming the two-time retarded Green's function to the real frequency axis using Eq.\eqref{eq:gretw}. In the non-interacting limit, the density of states for the two spin species are identical after the hopping parameter quench. We used a small Coulomb interaction $U=1.0$. The density of states for spin-$\uparrow$ and spin-$\downarrow$ exhibit very small differences at time $t=0^+$. We checked numerically that larger Coulomb interaction will induce a larger difference in the spin-resolved density of states. As time evolves, the density of states for the two spin species move toward each other and finally meet at $t=20$. By comparing the density of states for the thermalized state with effective temperature $T_{\mathrm{eff}}=0.5230$, we confirmed our conclusion in the previous paragraph.

The exponential decay with time of the order parameter depends on the Coulomb interaction, the initial mass imbalance ratio, and the initial temperature.  We will study the effect of one of the three factors by fixing the other two.  In Fig.\ref{fig:gqch_u}, we plot the order parameter of $SU(2)$ symmetry (deviation) as a function of time for different Coulomb interactions $U$ with fixed initial inverse temperature $\beta$ and initial mass imbalance ratio $r_0=V_\downarrow/V_\uparrow$. In the top panels, the order parameter $\gamma(t)$ (log scale $y$-axis) at different Coulomb interactions $U=0.5, 0.6,\cdots,1.0, 1.5, 2.0$ are plotted as a function of time. An approximate exponential decay is observed. The larger the Coulomb interaction, the faster the decay rate of the order parameter.
In the lower panels, the decay rate as a function of Coulomb interaction strength is plotted with open dots. 

The qualitative behaviors above can be understood from two limits. In the non-interacting limit, the momentum distribution after a hopping quench for different spins are $n_{\uparrow}(\epsilon) = 1/(e^{\beta\epsilon} +1)$ and $n_{\downarrow}(\epsilon) = 1/(e^{\beta\epsilon r_0} +1)$, respectively, where $r_0$ is the initial mass imbalance ratio. As time evolves, the momentum distribution does not change because the momentum is a good quantum number (no Coulomb scattering since $U=0$). This limit has an infinitely long approach time to the thermalized state ($SU(2)$ symmetry recovered). As a result, the decay rate is zero in the non-interacting limit.  In the infinite Coulomb interaction limit (atomic limit), the mass imbalance quench can be ignored (since the kinetic energy in either case is negligible) and so has no effect on the thermal distribution. In the infinite Coulomb interaction limit we can take the decay rate $\lambda = \infty$. Note, in the large Coulomb interaction region with relative low temperature, an antiferromagnetic state will appear. In this paper we restrict ourself to the weak interaction limit and relative high temperature to ensure we have a non-equilibrium paramagnetic solution. 
Further, based on second order perturbation theory (the first order terms will cancel due to particle-hole symmetry) a quadratic function of Coulomb interaction $\alpha U^2$ is plotted in the lower panels as a solid line, where $\alpha$ depends on the specific system parameters. The larger the Coulomb interaction, the greater the deviation from quadratic dependence of $U$ (higher order perturbation terms are needed). 

\begin{figure}[t]
\includegraphics[width=1.0\linewidth]{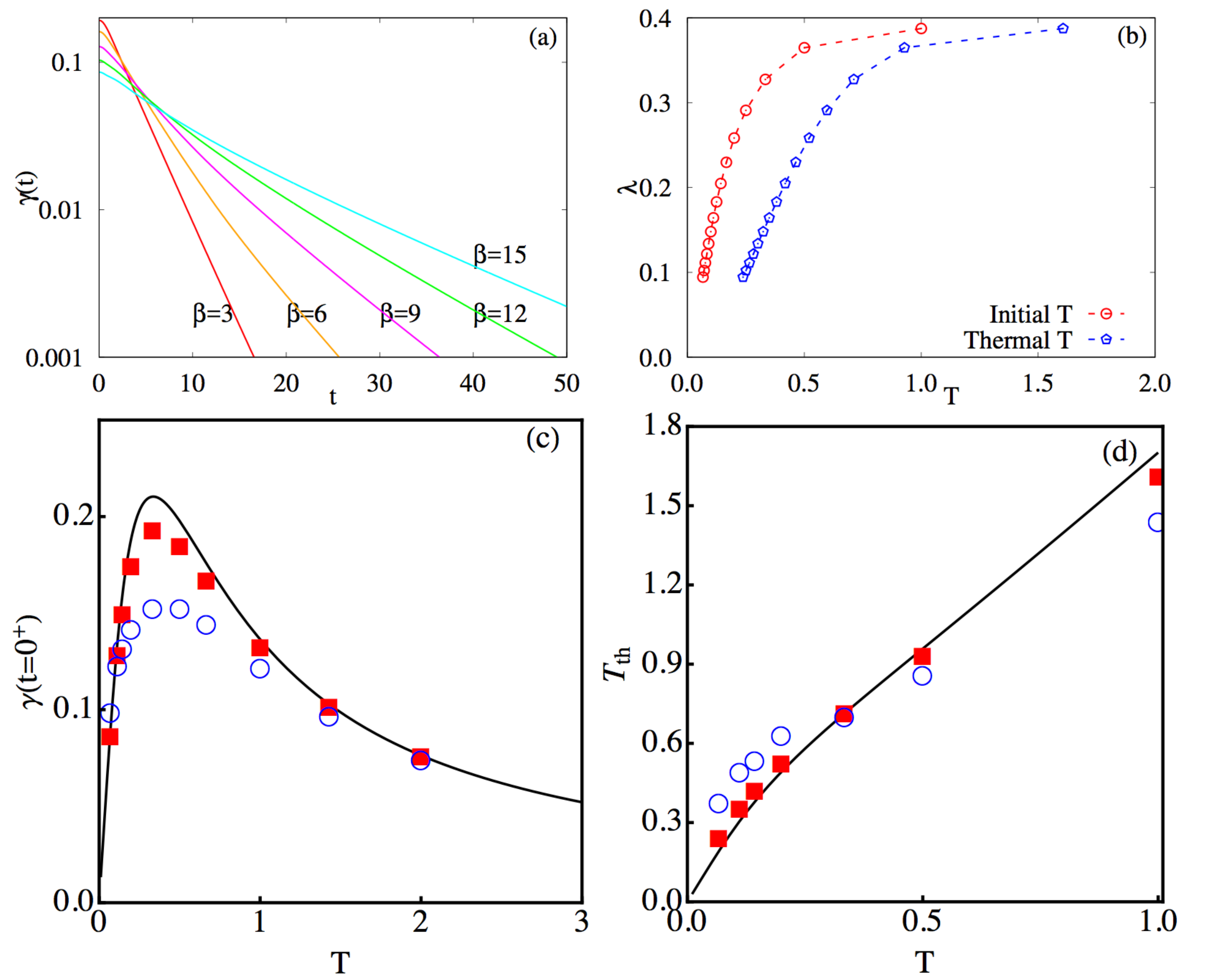}  %
\caption{(Color online) (a) Order parameter $\gamma(t)$ as a function of time for the half-filled mass imbalanced Hubbard model($V_\downarrow/V_\uparrow=1/4$) after a hopping integral quench $r=V_\downarrow/V_\uparrow$ from $r=1/4$ to $r=1$ at $U=1.0$. The inverse temperatures are shown from $\beta=3.0, 6.0, 9.0, 12.0, 15.0$. (b) The approximate decay rate as a function of initial temperature and thermalized temperature are plotted as circles and diamonds, respectively.
(c) Initial order parameters at $t=0^+$ are plotted as a function of initial temperature with filled squares for $U=1.0$ and open circles for $U=2.0$. As a comparison, the corresponding order parameter $\gamma(t=0^+)$ in the non-interacting limit $U=0.0$ is plotted as a solid line. (d) The thermalized temperature is plotted as a function of initial temperature with open cycles or $U=1.0$ and open circles for $U=2.0$.   As a comparison, the corresponding thermalized temperature $T_{th}$ for $U=0.0$ is plotted as a solid line. 
}
\label{fig:gqch_beta}
\end{figure}
\begin{figure}[t]
\includegraphics[width=1.0\linewidth]{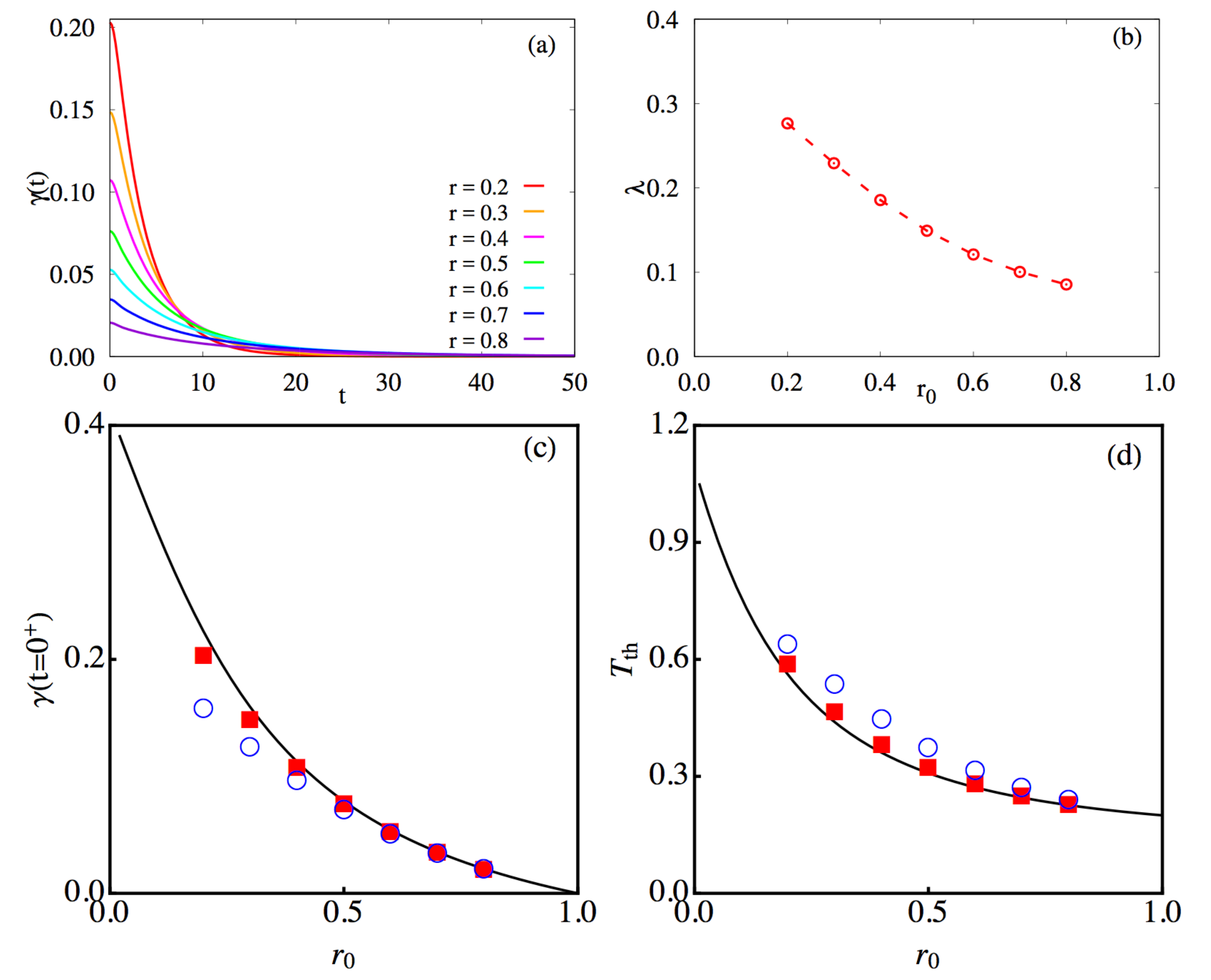}  %
\caption{(Color online) (a) Order parameter $\gamma(t)$ as a function of time for the half-filled mass imbalanced Hubbard model($V_\downarrow/V_\uparrow=1/4$) after an hopping integral quench $r=V_\downarrow/V_\uparrow$ from $r=0.2,0.3,0.4,\cdots,0.8$ to $r=1.0$ at $U=1.0$ and $\beta=5.0$ with different initial mass imbalance ratios (b) The approximate decay rate as a function of $r_0$. (c) The initial order parameter after the quench $\gamma(t=0^+)$ as a function of $r_0$ is plotted with open dots for $U=1.0$ and open circles for $U=2.0$. As a comparison, the corresponding order parameter $\gamma(t=0^+)$ with $U=0.0$ is plotted with a solid line.  (d) The thermalized temperature as a function $r_0$ is plotted with filled squares for $U=1.0$ and open circles for $U=2.0$.  As a comparison, the corresponding thermalized temperature for $U=0.0$ is plotted with a solid line}
\label{fig:gqch_r}
\end{figure}

By comparing Fig.\ref{fig:gqch_u} (b) with $r_0=1/4, \beta=5.0$ and (f) with $r_0=1/4, \beta=10.0$, we realized the decay rate in (f) is smaller than in (b) for each fixed Coulomb interaction. We conclude from this comparison that the decay rate depends on the initial temperature $1/\beta$ such that the lower temperature, the longer the time needed to relax to the thermalized state. A systematic study and discussion on the initial temperature dependence is illustrated in Fig.\ref{fig:gqch_beta}. By comparing Fig.\ref{fig:gqch_u} (b) with $r_0=1/4, \beta=5.0$ and (d) with $r_0=1/2, \beta=5.0$, one sees the decay rate in (d) is smaller than in (b) for each fixed Coulomb interaction. 
We conclude from this comparison that the decay rate depends on the initial mass imbalance ratio $r_0$ such that the larger mass imbalance ratio (closer to the final mass-balance Hubbard model Hamiltonian), longer the time that is needed to evolve to the thermalized state. A systematic study of the evolution dependence on the initial mass imbalance ratio is illustrated in Fig.\ref{fig:gqch_r}.  

By fixing the Coulomb interaction to be $U=1.0$ and the initial mass-imbalance ratio $r_0=1/4$, we plot the order parameter as a function of time at different temperatures in Fig.\ref{fig:gqch_beta}(a). 
At time $t=0^+$, one sees the order parameter is larger if the initial temperature $1/\beta$ is smaller. Since the Coulomb interaction is weak here, this behavior can be understood from the non-interacting limit. In the non-interacting limit, the order parameter after the hopping parameter quench from $V_{\downarrow}(t<0)=V_{\uparrow}r_0$ to $V_{\downarrow}(t>0)=V_{\uparrow}$ is given by,
\begin{align}
\gamma(t=0^+) = \int_{-2V_\uparrow}^{+2V_\uparrow} d\epsilon \rho_{\uparrow}(\epsilon) \left|\frac{1}{e^{\beta\epsilon} + 1} - \frac{1}{e^{\beta\epsilon r_0^{}} + 1} \right|,
\label{eq:gqch_u0}
\end{align}
where $r_0 = V_{\downarrow}(t<0) /V_{\uparrow}$ is the initial mass imbalance ratio. 

The order parameter $\gamma(t=0^+)$ as a function of temperature $T=1/\beta$ is plotted in Fig.\ref{fig:gqch_beta}(c) with a solid line for $U=0.0$. This plot can be understood physically from the zero temperature and infinite temperature limit. 
In the zero temperature limit the states with energy $\epsilon < 0$ ($\epsilon > 0$) are fully occupied (vacant). As one quenches the hopping integral of spin-$\downarrow$ electrons, the occupation at each $\epsilon$ is the same as before (occupied up to $\epsilon = 0$) and the order parameter will be zero. In the infinite temperature limit every state is equally populated and independent of its energy. 
The order parameter will be zero after hopping parameter quench.
In between these two limits, there exist a critical temperature $T \approx 0.3365 (\beta \approx 2.972)$ where order parameter increases (decreases) monotonically with the temperature below (above) the critical temperature. The initial order parameters for $U=1.0$ and $U=2.0$ are plotted with filled squares and open circles, respectively. One sees the biggest deviation from $U=0$ occurs at the critical temperature. Increasing the temperature beyond the critical one will decrease the deviation from the non-interacting limit. This can be explained as a competition between the kinetic energy and the Coulomb interaction. In high temperature region, the kinetic terms overcome the Coulomb terms and dominate the behavior of the order parameter. This picture in the high temperature region can be further confirmed by plotting the decay rate of the order parameter as a function of temperature in Fig.\ref{fig:gqch_beta}(b).  

First, we confirm the conclusion that in the low temperature regime, the decay rate increases monotonically with initial temperature. Physically, the relaxation process to the thermalized state is driven by the Coulomb scattering. 
Increasing the initial temperature will enhance the thermal fluctuations of electrons and enhance the collision probability leading to a larger decay rate. However, the decay rate tends to increase slower and finally saturates in the high temperature regime. 
When the temperate is high (kinetic energy overcomes the Coulomb energy), the states of the initial equilibrium state tend to converge, and the decay rates tend to saturate.
Since the temperature is only well-defined in the initial equilibrium states and the final thermalized states, we limit ourself to the qualitative analysis above. 

Finally, we plot the thermalized temperature as a function of initial temperature at $U=1.0$ in Fig.\ref{fig:gqch_beta}(d) with filled squares. A comparison with $U=0.0$ and $U=2.0$ are plotted with a solid line and open circles. A critical initial temperature is observed at $T \approx 0.33$, where Coulomb interaction tend to increase (decrease) the thermalized temperature comparing to the infinitesimal Coulomb interaction limit. 
 
By fixing the Coulomb interaction to be $U=1.0$ and initial temperature to be $\beta=5.0$, we plot the order parameter as a function of time for different initial mass imbalance ratios $r_0$ in Fig.\ref{fig:gqch_r}. 
In Fig.\ref{fig:gqch_r}(a), the order parameters is plotted as a function of time.  At time $t=0^+$, one sees the order parameter is larger if the mass-imbalance ratio is smaller. 
The initial order parameter $\gamma(t=0^+)$ as a function of $r_0$ is plotted in Fig.\ref{fig:gqch_r}(c) with filled squares. 
This can be understood again by considering the non-interacting limit in Eq.\eqref{eq:gqch_u0}. The order parameter $\gamma(t=0^+)$ at $U=0.0$ and $U=2.0$ are plotted in Fig.\ref{fig:gqch_r}(c) with a solid line and open circles, respectively. 
The order parameter decreases monotonically as $r_0$ is increased until the limit $r_0=1$ (order parameter is zero). The deviation from the non-interacting limit is larger as the initial mass imbalance ratio decreases.
Further, in Fig.\ref{fig:gqch_r}(b), the approximate decay rate is plotted as a function of the initial mass imbalance ratio $r_0$. 
Our results indicate that the decay rate decreases monotonically with increasing initial mass imbalance ratio. Finally, the thermalized temperature as a function of $r_0$ for $U=1.0$ is shown with filled squares. To illustrate the effect of Coulomb interaction, the data for $U=0.0$ and $U=2.0$ are plotted with a solid line and open circles, respectively. The thermalized temperature decreases as one increases the initial imbalance ratio.
In the limit $r_0=1$, the initial temperature equals the final thermalized temperature. 

\section{Dependence on the ramp shape and pulse form}
\label{sec:pulse}
\begin{figure}[t]
\includegraphics[width=1.0\linewidth]{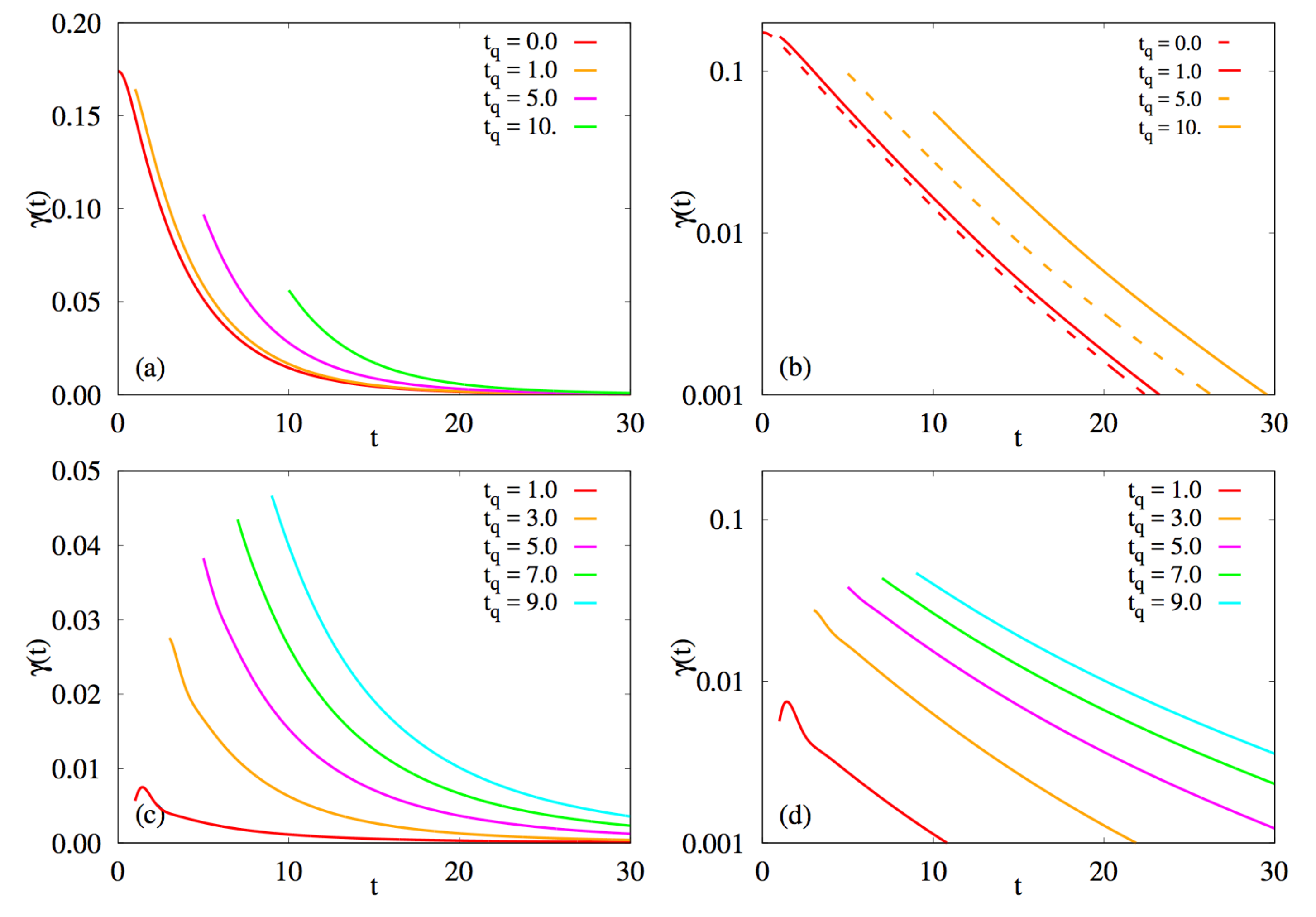}  %
\caption{(Color online) (a),(b) Order parameter $\gamma(t)$ as a function of time for the half-filled mass imbalanced Hubbard model after a hopping integral quench ramp $V_{\downarrow}(t\leq t_q) = V_{\downarrow}^i + (V_{\downarrow}^f-V_{\downarrow}^i)t/t_q$ at $U=1.0$.
(a) Normal scale for vertical axis at $\beta=5.0$. (b) Log scale for vertical axis at $\beta=5.0$.  (c),(d) Order parameter $\gamma(t)$ as a function of time for the half-filled mass balanced Hubbard model after a hopping integral pulse $V_{\downarrow}(t\leq t_q) = V_{\downarrow}^i + (V_{\downarrow}^f-V_{\downarrow}^i)\sin(\pi t/2 t_q)$ at $U=1.0$. (c) Normal scale for vertical axis axis.
(d) Log scale for vertical axis at $\beta=5.0$.
}
\label{fig:gqch_ramp}
\end{figure}

Experimentally, the change of parameters in the Hamiltonian takes a finite amount of time.  To model this, we suppose there exist a linear ramp to achieve the final parameter, 
\begin{equation}
V_{\downarrow}(t\leq t_q) = V_{\downarrow}^i + (V_{\downarrow}^f-V_{\downarrow}^i)t/t_q,
\end{equation}
where $t_q$ is the time used to achieve the final $SU(2)$ recovered Hamiltonian. In Fig.\ref{fig:gqch_ramp}(a-b), we plot the evolution of the order parameter as a function of time for different quench times $t_q=0.0,1.0,5.0,10.0$ at fixed Coulomb interaction $U=1.0$ and inverse temperature $\beta=5$. The decay rates for different linear ramps are approximately the same while a longer quench time leads to a longer relaxation time to a thermalized state. This is consistent with our previous study of the  decay rate dependence on the initial mass imbalance: In the linear ramp time ($0<t<t_q$), the ratio difference is smaller, so the decay rate in that time region is smaller. At time $t=t_q$, the order parameter will be larger than the quenched case ($t_q=0$).

Finally, we studied the case in which we have the mass balanced Hubbard model at time $t=0$ and apply a pulse change to the spin-$\downarrow$ hopping parameter change,
\begin{equation}
V_{\downarrow}(t\leq t_q) = V_{\downarrow}^i + (V_{\downarrow}^f-V_{\downarrow}^i)\sin(\pi t/2 t_q),
\end{equation}
with different quench time (width of pulse) $t_q = 1.0,3.0,5.0,7.0,9.0$. We plot the order parameter as a function of time in normal scale in Fig.\ref{fig:gqch_ramp}(c) and log scale in Fig.\ref{fig:gqch_ramp}(d).  As the quench time $t_q$ increases, the order parameter $\gamma(t_q)$ is larger.  Taken together, we see that the pulse shape can be used as a way to engineer the relaxation behavior of interacting, quantum many-particle systems.

\section{discussion and conclusion}
\label{sec:conclusion}
In this work, we theoretically studied the dynamical evolution towards $SU(2)$ symmetry of a system that is quenched from an $SU(2)$ broken one (mass-imbalanced Hubbard model) to an $SU(2)$ symmetry recovered one (mass-balanced Hubbard model). This model can be experimentally implemented in cold atom systems. We define the time dependent order parameter $\gamma(t)$ (total momentum-integrated difference between spin-$\uparrow$ and spin-$\downarrow$ momentum distribution) to characterize the $SU(2)$ symmetry. By comparing the spin-resolved momentum distribution of the $SU(2)$ symmetry recovered state (obtained for times such that $\gamma(t)=0$) with a thermalized state (an equilibrium state with effective temperature), we conclude the $SU(2)$ symmetry recovered state is a thermalized state. This conclusion is further confirmed by computing the spin- and time-resolved density of states.

Further, we observe the order parameter undergoes a nearly exponential decay towards the $SU(2)$ symmetry recovered states.  We studied the approximate decay rate and its relation to the initial temperature, Coulomb interaction strength, and the initial mass-imbalance ratio.  These dependences are studied by varying one parameter while fixing the other two. We found the order parameter in the weak Coulomb interaction region exhibits a nearly quadratic dependence on $U$, which can be interpreted with second order perturbation theory. For larger Coulomb interaction values, the deviation from quadratic dependence shows higher order terms must be taken into account.
 
We studied the dependence of approximate decay rate on the temperature. The decay rate increases rapidly with temperate in the low-temperature regime. By contrast, it saturates at higher temperatures (when the Coulomb interaction energy is overwhelmed relative to the kinetic energy).  We studied the initial order parameter after the quench $\gamma(t=0^+)$ and found a critical temperature where the order parameter increases (decreases) for temperatures below (above) the critical temperature.  The decay rate towards the thermalized state decreases as the initial imbalance ratio increases. Finally, we studied the dependence on the ramp shape and the pulse shape.  Taken together, our results provide a guide to engineer the relaxation behavior of interacting, quantum many-particle systems.

\section*{Acknowledgements} 
We acknowledge fruitful discussions with Qi Chen, Liang Dong, Quansheng Wu and Shengnan Zhang. We are grateful to Li Huang for a collaboration on a related project. L.D and G.A.F gratefully acknowledge funding from Army Research Office Grant No. W911NF- 14-1-0579, NSF Grant No. DMR-1507621, and NSF Materials Research Science and Engineering Center Grant No. DMR-1720595.  This work was performed in part at Aspen Center for Physics, which is supported by National Science Foundation grant PHY-1607611. 

\bibliography{tqch-massimbalance.bib}

\begin{thebibliography}{39}%
\makeatletter
\providecommand \@ifxundefined [1]{%
 \@ifx{#1\undefined}
}%
\providecommand \@ifnum [1]{%
 \ifnum #1\expandafter \@firstoftwo
 \else \expandafter \@secondoftwo
 \fi
}%
\providecommand \@ifx [1]{%
 \ifx #1\expandafter \@firstoftwo
 \else \expandafter \@secondoftwo
 \fi
}%
\providecommand \natexlab [1]{#1}%
\providecommand \enquote  [1]{``#1''}%
\providecommand \bibnamefont  [1]{#1}%
\providecommand \bibfnamefont [1]{#1}%
\providecommand \citenamefont [1]{#1}%
\providecommand \href@noop [0]{\@secondoftwo}%
\providecommand \href [0]{\begingroup \@sanitize@url \@href}%
\providecommand \@href[1]{\@@startlink{#1}\@@href}%
\providecommand \@@href[1]{\endgroup#1\@@endlink}%
\providecommand \@sanitize@url [0]{\catcode `\\12\catcode `\$12\catcode
  `\&12\catcode `\#12\catcode `\^12\catcode `\_12\catcode `\%12\relax}%
\providecommand \@@startlink[1]{}%
\providecommand \@@endlink[0]{}%
\providecommand \url  [0]{\begingroup\@sanitize@url \@url }%
\providecommand \@url [1]{\endgroup\@href {#1}{\urlprefix }}%
\providecommand \urlprefix  [0]{URL }%
\providecommand \Eprint [0]{\href }%
\providecommand \doibase [0]{http://dx.doi.org/}%
\providecommand \selectlanguage [0]{\@gobble}%
\providecommand \bibinfo  [0]{\@secondoftwo}%
\providecommand \bibfield  [0]{\@secondoftwo}%
\providecommand \translation [1]{[#1]}%
\providecommand \BibitemOpen [0]{}%
\providecommand \bibitemStop [0]{}%
\providecommand \bibitemNoStop [0]{.\EOS\space}%
\providecommand \EOS [0]{\spacefactor3000\relax}%
\providecommand \BibitemShut  [1]{\csname bibitem#1\endcsname}%
\let\auto@bib@innerbib\@empty
\bibitem [{\citenamefont {Lindner}\ \emph {et~al.}(2011)\citenamefont
  {Lindner}, \citenamefont {Refael},\ and\ \citenamefont
  {Galitski}}]{Lindner:nat11}%
  \BibitemOpen
  \bibfield  {author} {\bibinfo {author} {\bibfnamefont {N.~H.}\ \bibnamefont
  {Lindner}}, \bibinfo {author} {\bibfnamefont {G.}~\bibnamefont {Refael}}, \
  and\ \bibinfo {author} {\bibfnamefont {V.}~\bibnamefont {Galitski}},\ }\href
  {\doibase 10.1038/nphys1926} {\bibfield  {journal} {\bibinfo  {journal} {Nat.
  Phys.}\ }\textbf {\bibinfo {volume} {7}},\ \bibinfo {pages} {490} (\bibinfo
  {year} {2011})}\BibitemShut {NoStop}%
\bibitem [{\citenamefont {Gull}\ \emph {et~al.}(2011)\citenamefont {Gull},
  \citenamefont {Millis}, \citenamefont {Lichtenstein}, \citenamefont
  {Rubtsov}, \citenamefont {Troyer},\ and\ \citenamefont
  {Werner}}]{Gull:rmp11}%
  \BibitemOpen
  \bibfield  {author} {\bibinfo {author} {\bibfnamefont {E.}~\bibnamefont
  {Gull}}, \bibinfo {author} {\bibfnamefont {A.~J.}\ \bibnamefont {Millis}},
  \bibinfo {author} {\bibfnamefont {A.~I.}\ \bibnamefont {Lichtenstein}},
  \bibinfo {author} {\bibfnamefont {A.~N.}\ \bibnamefont {Rubtsov}}, \bibinfo
  {author} {\bibfnamefont {M.}~\bibnamefont {Troyer}}, \ and\ \bibinfo {author}
  {\bibfnamefont {P.}~\bibnamefont {Werner}},\ }\href {\doibase
  10.1103/RevModPhys.83.349} {\bibfield  {journal} {\bibinfo  {journal} {Rev.
  Mod. Phys.}\ }\textbf {\bibinfo {volume} {83}},\ \bibinfo {pages} {349}
  (\bibinfo {year} {2011})}\BibitemShut {NoStop}%
\bibitem [{\citenamefont {Wang}\ \emph {et~al.}(2013)\citenamefont {Wang},
  \citenamefont {Steinberg}, \citenamefont {Jarillo-Herrero},\ and\
  \citenamefont {Gedik}}]{Wang:sci13}%
  \BibitemOpen
  \bibfield  {author} {\bibinfo {author} {\bibfnamefont {Y.~H.}\ \bibnamefont
  {Wang}}, \bibinfo {author} {\bibfnamefont {H.}~\bibnamefont {Steinberg}},
  \bibinfo {author} {\bibfnamefont {P.}~\bibnamefont {Jarillo-Herrero}}, \ and\
  \bibinfo {author} {\bibfnamefont {N.}~\bibnamefont {Gedik}},\ }\href
  {\doibase 10.1126/science.1239834} {\bibfield  {journal} {\bibinfo  {journal}
  {Science}\ }\textbf {\bibinfo {volume} {342}},\ \bibinfo {pages} {453}
  (\bibinfo {year} {2013})}\BibitemShut {NoStop}%
\bibitem [{\citenamefont {Aoki}\ \emph {et~al.}(2014)\citenamefont {Aoki},
  \citenamefont {Tsuji}, \citenamefont {Eckstein}, \citenamefont {Kollar},
  \citenamefont {Oka},\ and\ \citenamefont {Werner}}]{Aoki:rmp14}%
  \BibitemOpen
  \bibfield  {author} {\bibinfo {author} {\bibfnamefont {H.}~\bibnamefont
  {Aoki}}, \bibinfo {author} {\bibfnamefont {N.}~\bibnamefont {Tsuji}},
  \bibinfo {author} {\bibfnamefont {M.}~\bibnamefont {Eckstein}}, \bibinfo
  {author} {\bibfnamefont {M.}~\bibnamefont {Kollar}}, \bibinfo {author}
  {\bibfnamefont {T.}~\bibnamefont {Oka}}, \ and\ \bibinfo {author}
  {\bibfnamefont {P.}~\bibnamefont {Werner}},\ }\href {\doibase
  10.1103/RevModPhys.86.779} {\bibfield  {journal} {\bibinfo  {journal} {Rev.
  Mod. Phys.}\ }\textbf {\bibinfo {volume} {86}},\ \bibinfo {pages} {779}
  (\bibinfo {year} {2014})}\BibitemShut {NoStop}%
\bibitem [{\citenamefont {Manmana}\ \emph {et~al.}(2007)\citenamefont
  {Manmana}, \citenamefont {Wessel}, \citenamefont {Noack},\ and\ \citenamefont
  {Muramatsu}}]{Manmana:prl07}%
  \BibitemOpen
  \bibfield  {author} {\bibinfo {author} {\bibfnamefont {S.~R.}\ \bibnamefont
  {Manmana}}, \bibinfo {author} {\bibfnamefont {S.}~\bibnamefont {Wessel}},
  \bibinfo {author} {\bibfnamefont {R.~M.}\ \bibnamefont {Noack}}, \ and\
  \bibinfo {author} {\bibfnamefont {A.}~\bibnamefont {Muramatsu}},\ }\href
  {\doibase 10.1103/PhysRevLett.98.210405} {\bibfield  {journal} {\bibinfo
  {journal} {Phys. Rev. Lett.}\ }\textbf {\bibinfo {volume} {98}},\ \bibinfo
  {pages} {210405} (\bibinfo {year} {2007})}\BibitemShut {NoStop}%
\bibitem [{\citenamefont {Rigol}\ \emph {et~al.}(2007)\citenamefont {Rigol},
  \citenamefont {Dunjko}, \citenamefont {Yurovsky},\ and\ \citenamefont
  {Olshanii}}]{Rigol:prl07}%
  \BibitemOpen
  \bibfield  {author} {\bibinfo {author} {\bibfnamefont {M.}~\bibnamefont
  {Rigol}}, \bibinfo {author} {\bibfnamefont {V.}~\bibnamefont {Dunjko}},
  \bibinfo {author} {\bibfnamefont {V.}~\bibnamefont {Yurovsky}}, \ and\
  \bibinfo {author} {\bibfnamefont {M.}~\bibnamefont {Olshanii}},\ }\href
  {\doibase 10.1103/PhysRevLett.98.050405} {\bibfield  {journal} {\bibinfo
  {journal} {Phys. Rev. Lett.}\ }\textbf {\bibinfo {volume} {98}},\ \bibinfo
  {pages} {050405} (\bibinfo {year} {2007})}\BibitemShut {NoStop}%
\bibitem [{\citenamefont {Kollath}\ \emph {et~al.}(2007)\citenamefont
  {Kollath}, \citenamefont {L{\"{a}}uchli},\ and\ \citenamefont
  {Altman}}]{Kollath:prb07}%
  \BibitemOpen
  \bibfield  {author} {\bibinfo {author} {\bibfnamefont {C.}~\bibnamefont
  {Kollath}}, \bibinfo {author} {\bibfnamefont {A.~M.}\ \bibnamefont
  {L{\"{a}}uchli}}, \ and\ \bibinfo {author} {\bibfnamefont {E.}~\bibnamefont
  {Altman}},\ }\href {\doibase 10.1103/PhysRevLett.98.180601} {\bibfield
  {journal} {\bibinfo  {journal} {Phys. Rev. Lett.}\ }\textbf {\bibinfo
  {volume} {98}},\ \bibinfo {pages} {180601} (\bibinfo {year}
  {2007})}\BibitemShut {NoStop}%
\bibitem [{\citenamefont {Mentink}\ and\ \citenamefont
  {Eckstein}(2014)}]{Mentink:prl14}%
  \BibitemOpen
  \bibfield  {author} {\bibinfo {author} {\bibfnamefont {J.~H.}\ \bibnamefont
  {Mentink}}\ and\ \bibinfo {author} {\bibfnamefont {M.}~\bibnamefont
  {Eckstein}},\ }\href {\doibase 10.1103/PhysRevLett.113.057201} {\bibfield
  {journal} {\bibinfo  {journal} {Phys. Rev. Lett.}\ }\textbf {\bibinfo
  {volume} {113}},\ \bibinfo {pages} {057201} (\bibinfo {year}
  {2014})}\BibitemShut {NoStop}%
\bibitem [{\citenamefont {Mentink}\ \emph {et~al.}(2015)\citenamefont
  {Mentink}, \citenamefont {Balzer},\ and\ \citenamefont
  {Eckstein}}]{Mentink:nc15}%
  \BibitemOpen
  \bibfield  {author} {\bibinfo {author} {\bibfnamefont {J.~H.}\ \bibnamefont
  {Mentink}}, \bibinfo {author} {\bibfnamefont {K.}~\bibnamefont {Balzer}}, \
  and\ \bibinfo {author} {\bibfnamefont {M.}~\bibnamefont {Eckstein}},\ }\href
  {\doibase 10.1038/ncomms7708} {\bibfield  {journal} {\bibinfo  {journal}
  {Nat. Commun.}\ }\textbf {\bibinfo {volume} {6}},\ \bibinfo {pages} {6708}
  (\bibinfo {year} {2015})}\BibitemShut {NoStop}%
\bibitem [{\citenamefont {Eckstein}\ and\ \citenamefont
  {Werner}(2016)}]{Eckstein:scirep16}%
  \BibitemOpen
  \bibfield  {author} {\bibinfo {author} {\bibfnamefont {M.}~\bibnamefont
  {Eckstein}}\ and\ \bibinfo {author} {\bibfnamefont {P.}~\bibnamefont
  {Werner}},\ }\href {\doibase 10.1038/srep21235} {\bibfield  {journal}
  {\bibinfo  {journal} {Sci. Rep.}\ }\textbf {\bibinfo {volume} {6}},\ \bibinfo
  {pages} {21235} (\bibinfo {year} {2016})}\BibitemShut {NoStop}%
\bibitem [{\citenamefont {Bukov}\ \emph {et~al.}(2016)\citenamefont {Bukov},
  \citenamefont {Kolodrubetz},\ and\ \citenamefont
  {Polkovnikov}}]{Bukov:prl16}%
  \BibitemOpen
  \bibfield  {author} {\bibinfo {author} {\bibfnamefont {M.}~\bibnamefont
  {Bukov}}, \bibinfo {author} {\bibfnamefont {M.}~\bibnamefont {Kolodrubetz}},
  \ and\ \bibinfo {author} {\bibfnamefont {A.}~\bibnamefont {Polkovnikov}},\
  }\href {\doibase 10.1103/PhysRevLett.116.125301} {\bibfield  {journal}
  {\bibinfo  {journal} {Phys. Rev. Lett.}\ }\textbf {\bibinfo {volume} {116}},\
  \bibinfo {pages} {125301} (\bibinfo {year} {2016})}\BibitemShut {NoStop}%
\bibitem [{\citenamefont {Werner}\ \emph {et~al.}(2012)\citenamefont {Werner},
  \citenamefont {Tsuji},\ and\ \citenamefont {Eckstein}}]{Werner:prb12}%
  \BibitemOpen
  \bibfield  {author} {\bibinfo {author} {\bibfnamefont {P.}~\bibnamefont
  {Werner}}, \bibinfo {author} {\bibfnamefont {N.}~\bibnamefont {Tsuji}}, \
  and\ \bibinfo {author} {\bibfnamefont {M.}~\bibnamefont {Eckstein}},\ }\href
  {\doibase 10.1103/PhysRevB.86.205101} {\bibfield  {journal} {\bibinfo
  {journal} {Phys. Rev. B}\ }\textbf {\bibinfo {volume} {86}},\ \bibinfo
  {pages} {205101} (\bibinfo {year} {2012})}\BibitemShut {NoStop}%
\bibitem [{\citenamefont {Tsuji}\ \emph {et~al.}(2013)\citenamefont {Tsuji},
  \citenamefont {Eckstein},\ and\ \citenamefont {Werner}}]{Tsuji:prl13}%
  \BibitemOpen
  \bibfield  {author} {\bibinfo {author} {\bibfnamefont {N.}~\bibnamefont
  {Tsuji}}, \bibinfo {author} {\bibfnamefont {M.}~\bibnamefont {Eckstein}}, \
  and\ \bibinfo {author} {\bibfnamefont {P.}~\bibnamefont {Werner}},\ }\href
  {\doibase 10.1103/PhysRevLett.110.136404} {\bibfield  {journal} {\bibinfo
  {journal} {Phys. Rev. Lett.}\ }\textbf {\bibinfo {volume} {110}},\ \bibinfo
  {pages} {136404} (\bibinfo {year} {2013})}\BibitemShut {NoStop}%
\bibitem [{\citenamefont {Tsuji}\ and\ \citenamefont
  {Werner}(2013)}]{Tsuji:prb13}%
  \BibitemOpen
  \bibfield  {author} {\bibinfo {author} {\bibfnamefont {N.}~\bibnamefont
  {Tsuji}}\ and\ \bibinfo {author} {\bibfnamefont {P.}~\bibnamefont {Werner}},\
  }\href {\doibase 10.1103/PhysRevB.88.165115} {\bibfield  {journal} {\bibinfo
  {journal} {Phys. Rev. B}\ }\textbf {\bibinfo {volume} {88}},\ \bibinfo
  {pages} {165115} (\bibinfo {year} {2013})}\BibitemShut {NoStop}%
\bibitem [{\citenamefont {Oka}\ and\ \citenamefont {Aoki}(2009)}]{Oka:prb09}%
  \BibitemOpen
  \bibfield  {author} {\bibinfo {author} {\bibfnamefont {T.}~\bibnamefont
  {Oka}}\ and\ \bibinfo {author} {\bibfnamefont {H.}~\bibnamefont {Aoki}},\
  }\href {http://link.aps.org/doi/10.1103/PhysRevB.79.081406} {\bibfield
  {journal} {\bibinfo  {journal} {Phys. Rev. B}\ }\textbf {\bibinfo {volume}
  {79}},\ \bibinfo {pages} {081406} (\bibinfo {year} {2009})}\BibitemShut
  {NoStop}%
\bibitem [{\citenamefont {Du}\ \emph {et~al.}(2017{\natexlab{a}})\citenamefont
  {Du}, \citenamefont {Zhou},\ and\ \citenamefont {Fiete}}]{Du:prb17a}%
  \BibitemOpen
  \bibfield  {author} {\bibinfo {author} {\bibfnamefont {L.}~\bibnamefont
  {Du}}, \bibinfo {author} {\bibfnamefont {X.}~\bibnamefont {Zhou}}, \ and\
  \bibinfo {author} {\bibfnamefont {G.~A.}\ \bibnamefont {Fiete}},\ }\href
  {\doibase 10.1103/PhysRevB.95.035136} {\bibfield  {journal} {\bibinfo
  {journal} {Phys. Rev. B}\ }\textbf {\bibinfo {volume} {95}},\ \bibinfo
  {pages} {035136} (\bibinfo {year} {2017}{\natexlab{a}})}\BibitemShut
  {NoStop}%
\bibitem [{\citenamefont {Du}\ and\ \citenamefont {Fiete}(2017)}]{Du:prb17b}%
  \BibitemOpen
  \bibfield  {author} {\bibinfo {author} {\bibfnamefont {L.}~\bibnamefont
  {Du}}\ and\ \bibinfo {author} {\bibfnamefont {G.~A.}\ \bibnamefont {Fiete}},\
  }\href {\doibase 10.1103/PhysRevB.95.235309} {\bibfield  {journal} {\bibinfo
  {journal} {Phys. Rev. B}\ }\textbf {\bibinfo {volume} {95}},\ \bibinfo
  {pages} {235309} (\bibinfo {year} {2017})}\BibitemShut {NoStop}%
\bibitem [{\citenamefont {Georges}\ \emph {et~al.}(1996)\citenamefont
  {Georges}, \citenamefont {Kotliar}, \citenamefont {Krauth},\ and\
  \citenamefont {Rozenberg}}]{Georges:rmp96}%
  \BibitemOpen
  \bibfield  {author} {\bibinfo {author} {\bibfnamefont {A.}~\bibnamefont
  {Georges}}, \bibinfo {author} {\bibfnamefont {G.}~\bibnamefont {Kotliar}},
  \bibinfo {author} {\bibfnamefont {W.}~\bibnamefont {Krauth}}, \ and\ \bibinfo
  {author} {\bibfnamefont {M.~J.}\ \bibnamefont {Rozenberg}},\ }\href {\doibase
  10.1103/RevModPhys.68.13} {\bibfield  {journal} {\bibinfo  {journal} {Rev.
  Mod. Phys.}\ }\textbf {\bibinfo {volume} {68}},\ \bibinfo {pages} {13}
  (\bibinfo {year} {1996})}\BibitemShut {NoStop}%
\bibitem [{\citenamefont {Wolf}\ \emph {et~al.}(2014)\citenamefont {Wolf},
  \citenamefont {McCulloch},\ and\ \citenamefont
  {Schollw{\"{o}}ck}}]{Wolf:prb14}%
  \BibitemOpen
  \bibfield  {author} {\bibinfo {author} {\bibfnamefont {F.~A.}\ \bibnamefont
  {Wolf}}, \bibinfo {author} {\bibfnamefont {I.~P.}\ \bibnamefont {McCulloch}},
  \ and\ \bibinfo {author} {\bibfnamefont {U.}~\bibnamefont
  {Schollw{\"{o}}ck}},\ }\href {\doibase 10.1103/PhysRevB.90.235131} {\bibfield
   {journal} {\bibinfo  {journal} {Phys. Rev. B}\ }\textbf {\bibinfo {volume}
  {90}},\ \bibinfo {pages} {235131} (\bibinfo {year} {2014})}\BibitemShut
  {NoStop}%
\bibitem [{\citenamefont {Balzer}\ \emph {et~al.}(2015)\citenamefont {Balzer},
  \citenamefont {Li}, \citenamefont {Vendrell},\ and\ \citenamefont
  {Eckstein}}]{Balzer:prb15}%
  \BibitemOpen
  \bibfield  {author} {\bibinfo {author} {\bibfnamefont {K.}~\bibnamefont
  {Balzer}}, \bibinfo {author} {\bibfnamefont {Z.}~\bibnamefont {Li}}, \bibinfo
  {author} {\bibfnamefont {O.}~\bibnamefont {Vendrell}}, \ and\ \bibinfo
  {author} {\bibfnamefont {M.}~\bibnamefont {Eckstein}},\ }\href {\doibase
  10.1103/PhysRevB.91.045136} {\bibfield  {journal} {\bibinfo  {journal} {Phys.
  Rev. B}\ }\textbf {\bibinfo {volume} {91}},\ \bibinfo {pages} {045136}
  (\bibinfo {year} {2015})}\BibitemShut {NoStop}%
\bibitem [{\citenamefont {Cohen}\ \emph {et~al.}(2015)\citenamefont {Cohen},
  \citenamefont {Gull}, \citenamefont {Reichman},\ and\ \citenamefont
  {Millis}}]{Cohen:prl15}%
  \BibitemOpen
  \bibfield  {author} {\bibinfo {author} {\bibfnamefont {G.}~\bibnamefont
  {Cohen}}, \bibinfo {author} {\bibfnamefont {E.}~\bibnamefont {Gull}},
  \bibinfo {author} {\bibfnamefont {D.~R.}\ \bibnamefont {Reichman}}, \ and\
  \bibinfo {author} {\bibfnamefont {A.~J.}\ \bibnamefont {Millis}},\ }\href
  {\doibase 10.1103/PhysRevLett.115.266802} {\bibfield  {journal} {\bibinfo
  {journal} {Phys. Rev. Lett.}\ }\textbf {\bibinfo {volume} {115}},\ \bibinfo
  {pages} {266802} (\bibinfo {year} {2015})}\BibitemShut {NoStop}%
\bibitem [{\citenamefont {Dong}\ \emph {et~al.}(2017)\citenamefont {Dong},
  \citenamefont {Krivenko}, \citenamefont {Kleinhenz}, \citenamefont {Antipov},
  \citenamefont {Cohen},\ and\ \citenamefont {Gull}}]{Dong:prb17}%
  \BibitemOpen
  \bibfield  {author} {\bibinfo {author} {\bibfnamefont {Q.}~\bibnamefont
  {Dong}}, \bibinfo {author} {\bibfnamefont {I.}~\bibnamefont {Krivenko}},
  \bibinfo {author} {\bibfnamefont {J.}~\bibnamefont {Kleinhenz}}, \bibinfo
  {author} {\bibfnamefont {A.~E.}\ \bibnamefont {Antipov}}, \bibinfo {author}
  {\bibfnamefont {G.}~\bibnamefont {Cohen}}, \ and\ \bibinfo {author}
  {\bibfnamefont {E.}~\bibnamefont {Gull}},\ }\href {\doibase
  10.1103/PhysRevB.96.155126} {\bibfield  {journal} {\bibinfo  {journal} {Phys.
  Rev. B}\ }\textbf {\bibinfo {volume} {96}},\ \bibinfo {pages} {155126}
  (\bibinfo {year} {2017})},\ \Eprint {http://arxiv.org/abs/1706.02975}
  {arXiv:1706.02975} \BibitemShut {NoStop}%
\bibitem [{\citenamefont {Eckstein}\ and\ \citenamefont
  {Werner}(2010)}]{Eckstein:prb10b}%
  \BibitemOpen
  \bibfield  {author} {\bibinfo {author} {\bibfnamefont {M.}~\bibnamefont
  {Eckstein}}\ and\ \bibinfo {author} {\bibfnamefont {P.}~\bibnamefont
  {Werner}},\ }\href {\doibase 10.1103/PhysRevB.82.115115} {\bibfield
  {journal} {\bibinfo  {journal} {Phys. Rev. B}\ }\textbf {\bibinfo {volume}
  {82}},\ \bibinfo {pages} {115115} (\bibinfo {year} {2010})}\BibitemShut
  {NoStop}%
\bibitem [{\citenamefont {Freericks}\ \emph {et~al.}(2006)\citenamefont
  {Freericks}, \citenamefont {Turkowski},\ and\ \citenamefont
  {Zlati{\'{c}}}}]{Freericks:prl06}%
  \BibitemOpen
  \bibfield  {author} {\bibinfo {author} {\bibfnamefont {J.~K.}\ \bibnamefont
  {Freericks}}, \bibinfo {author} {\bibfnamefont {V.~M.}\ \bibnamefont
  {Turkowski}}, \ and\ \bibinfo {author} {\bibfnamefont {V.}~\bibnamefont
  {Zlati{\'{c}}}},\ }\href {\doibase 10.1103/PhysRevLett.97.266408} {\bibfield
  {journal} {\bibinfo  {journal} {Phys. Rev. Lett.}\ }\textbf {\bibinfo
  {volume} {97}},\ \bibinfo {pages} {266408} (\bibinfo {year}
  {2006})}\BibitemShut {NoStop}%
\bibitem [{\citenamefont {Eckstein}\ and\ \citenamefont
  {Kollar}(2008)}]{Eckstein:prl08}%
  \BibitemOpen
  \bibfield  {author} {\bibinfo {author} {\bibfnamefont {M.}~\bibnamefont
  {Eckstein}}\ and\ \bibinfo {author} {\bibfnamefont {M.}~\bibnamefont
  {Kollar}},\ }\href {\doibase 10.1103/PhysRevLett.100.120404} {\bibfield
  {journal} {\bibinfo  {journal} {Phys. Rev. Lett.}\ }\textbf {\bibinfo
  {volume} {100}},\ \bibinfo {pages} {120404} (\bibinfo {year}
  {2008})}\BibitemShut {NoStop}%
\bibitem [{\citenamefont {Freericks}\ and\ \citenamefont
  {Zlati{\'{c}}}(2003)}]{Freericks:rmp03}%
  \BibitemOpen
  \bibfield  {author} {\bibinfo {author} {\bibfnamefont {J.}~\bibnamefont
  {Freericks}}\ and\ \bibinfo {author} {\bibfnamefont {V.}~\bibnamefont
  {Zlati{\'{c}}}},\ }\href {\doibase 10.1103/RevModPhys.75.1333} {\bibfield
  {journal} {\bibinfo  {journal} {Rev. Mod. Phys.}\ }\textbf {\bibinfo {volume}
  {75}},\ \bibinfo {pages} {1333} (\bibinfo {year} {2003})}\BibitemShut
  {NoStop}%
\bibitem [{\citenamefont {Eckstein}\ \emph {et~al.}(2009)\citenamefont
  {Eckstein}, \citenamefont {Kollar},\ and\ \citenamefont
  {Werner}}]{Eckstein:prl09}%
  \BibitemOpen
  \bibfield  {author} {\bibinfo {author} {\bibfnamefont {M.}~\bibnamefont
  {Eckstein}}, \bibinfo {author} {\bibfnamefont {M.}~\bibnamefont {Kollar}}, \
  and\ \bibinfo {author} {\bibfnamefont {P.}~\bibnamefont {Werner}},\ }\href
  {\doibase 10.1103/PhysRevLett.103.056403} {\bibfield  {journal} {\bibinfo
  {journal} {Phys. Rev. Lett.}\ }\textbf {\bibinfo {volume} {103}},\ \bibinfo
  {pages} {056403} (\bibinfo {year} {2009})}\BibitemShut {NoStop}%
\bibitem [{\citenamefont {Eckstein}\ \emph {et~al.}(2010)\citenamefont
  {Eckstein}, \citenamefont {Kollar},\ and\ \citenamefont
  {Werner}}]{Eckstein:prb10a}%
  \BibitemOpen
  \bibfield  {author} {\bibinfo {author} {\bibfnamefont {M.}~\bibnamefont
  {Eckstein}}, \bibinfo {author} {\bibfnamefont {M.}~\bibnamefont {Kollar}}, \
  and\ \bibinfo {author} {\bibfnamefont {P.}~\bibnamefont {Werner}},\ }\href
  {\doibase 10.1103/PhysRevB.81.115131} {\bibfield  {journal} {\bibinfo
  {journal} {Phys. Rev. B}\ }\textbf {\bibinfo {volume} {81}},\ \bibinfo
  {pages} {115131} (\bibinfo {year} {2010})}\BibitemShut {NoStop}%
\bibitem [{\citenamefont {Freericks}(2008)}]{Freericks:prb08}%
  \BibitemOpen
  \bibfield  {author} {\bibinfo {author} {\bibfnamefont {J.}~\bibnamefont
  {Freericks}},\ }\href {\doibase 10.1103/PhysRevB.77.075109} {\bibfield
  {journal} {\bibinfo  {journal} {Phys. Rev. B}\ }\textbf {\bibinfo {volume}
  {77}},\ \bibinfo {pages} {075109} (\bibinfo {year} {2008})}\BibitemShut
  {NoStop}%
\bibitem [{\citenamefont {Eckstein}\ and\ \citenamefont
  {Werner}(2011)}]{Eckstein:prl11}%
  \BibitemOpen
  \bibfield  {author} {\bibinfo {author} {\bibfnamefont {M.}~\bibnamefont
  {Eckstein}}\ and\ \bibinfo {author} {\bibfnamefont {P.}~\bibnamefont
  {Werner}},\ }\href {\doibase 10.1103/PhysRevLett.107.186406} {\bibfield
  {journal} {\bibinfo  {journal} {Phys. Rev. Lett.}\ }\textbf {\bibinfo
  {volume} {107}},\ \bibinfo {pages} {186406} (\bibinfo {year}
  {2011})}\BibitemShut {NoStop}%
\bibitem [{\citenamefont {Fotso}\ \emph {et~al.}(2015)\citenamefont {Fotso},
  \citenamefont {Mikelsons},\ and\ \citenamefont {Freericks}}]{Fotso:scirep14}%
  \BibitemOpen
  \bibfield  {author} {\bibinfo {author} {\bibfnamefont {H.}~\bibnamefont
  {Fotso}}, \bibinfo {author} {\bibfnamefont {K.}~\bibnamefont {Mikelsons}}, \
  and\ \bibinfo {author} {\bibfnamefont {J.~K.}\ \bibnamefont {Freericks}},\
  }\href {\doibase 10.1038/srep04699} {\bibfield  {journal} {\bibinfo
  {journal} {Sci. Rep.}\ }\textbf {\bibinfo {volume} {4}},\ \bibinfo {pages}
  {4699} (\bibinfo {year} {2015})}\BibitemShut {NoStop}%
\bibitem [{\citenamefont {Tsuji}\ \emph {et~al.}(2011)\citenamefont {Tsuji},
  \citenamefont {Oka}, \citenamefont {Werner},\ and\ \citenamefont
  {Aoki}}]{Tsuji:prl11}%
  \BibitemOpen
  \bibfield  {author} {\bibinfo {author} {\bibfnamefont {N.}~\bibnamefont
  {Tsuji}}, \bibinfo {author} {\bibfnamefont {T.}~\bibnamefont {Oka}}, \bibinfo
  {author} {\bibfnamefont {P.}~\bibnamefont {Werner}}, \ and\ \bibinfo {author}
  {\bibfnamefont {H.}~\bibnamefont {Aoki}},\ }\href {\doibase
  10.1103/PhysRevLett.106.236401} {\bibfield  {journal} {\bibinfo  {journal}
  {Phys. Rev. Lett.}\ }\textbf {\bibinfo {volume} {106}},\ \bibinfo {pages}
  {236401} (\bibinfo {year} {2011})}\BibitemShut {NoStop}%
\bibitem [{\citenamefont {Nguyen}\ \emph {et~al.}(2015)\citenamefont {Nguyen},
  \citenamefont {Phung}, \citenamefont {Phan},\ and\ \citenamefont
  {Tran}}]{Nguyen:prb15}%
  \BibitemOpen
  \bibfield  {author} {\bibinfo {author} {\bibfnamefont {D.-B.}\ \bibnamefont
  {Nguyen}}, \bibinfo {author} {\bibfnamefont {D.-K.}\ \bibnamefont {Phung}},
  \bibinfo {author} {\bibfnamefont {V.-N.}\ \bibnamefont {Phan}}, \ and\
  \bibinfo {author} {\bibfnamefont {M.-T.}\ \bibnamefont {Tran}},\ }\href
  {\doibase 10.1103/PhysRevB.91.115140} {\bibfield  {journal} {\bibinfo
  {journal} {Phys. Rev. B}\ }\textbf {\bibinfo {volume} {91}},\ \bibinfo
  {pages} {115140} (\bibinfo {year} {2015})}\BibitemShut {NoStop}%
\bibitem [{\citenamefont {Dao}\ \emph {et~al.}(2012)\citenamefont {Dao},
  \citenamefont {Ferrero}, \citenamefont {Cornaglia},\ and\ \citenamefont
  {Capone}}]{Dao:pra12}%
  \BibitemOpen
  \bibfield  {author} {\bibinfo {author} {\bibfnamefont {T.-L.}\ \bibnamefont
  {Dao}}, \bibinfo {author} {\bibfnamefont {M.}~\bibnamefont {Ferrero}},
  \bibinfo {author} {\bibfnamefont {P.~S.}\ \bibnamefont {Cornaglia}}, \ and\
  \bibinfo {author} {\bibfnamefont {M.}~\bibnamefont {Capone}},\ }\href
  {\doibase 10.1103/PhysRevA.85.013606} {\bibfield  {journal} {\bibinfo
  {journal} {Phys. Rev. A}\ }\textbf {\bibinfo {volume} {85}},\ \bibinfo
  {pages} {013606} (\bibinfo {year} {2012})}\BibitemShut {NoStop}%
\bibitem [{\citenamefont {Liu}\ and\ \citenamefont {Wang}(2015)}]{Liu:prb15}%
  \BibitemOpen
  \bibfield  {author} {\bibinfo {author} {\bibfnamefont {Y.-H.}\ \bibnamefont
  {Liu}}\ and\ \bibinfo {author} {\bibfnamefont {L.}~\bibnamefont {Wang}},\
  }\href {\doibase 10.1103/PhysRevB.92.235129} {\bibfield  {journal} {\bibinfo
  {journal} {Phys. Rev. B}\ }\textbf {\bibinfo {volume} {92}},\ \bibinfo
  {pages} {235129} (\bibinfo {year} {2015})}\BibitemShut {NoStop}%
\bibitem [{\citenamefont {Philipp}\ \emph {et~al.}(2017)\citenamefont
  {Philipp}, \citenamefont {Wallerberger}, \citenamefont {Gunacker},\ and\
  \citenamefont {Held}}]{Philipp:epjb17}%
  \BibitemOpen
  \bibfield  {author} {\bibinfo {author} {\bibfnamefont {M.-T.}\ \bibnamefont
  {Philipp}}, \bibinfo {author} {\bibfnamefont {M.}~\bibnamefont
  {Wallerberger}}, \bibinfo {author} {\bibfnamefont {P.}~\bibnamefont
  {Gunacker}}, \ and\ \bibinfo {author} {\bibfnamefont {K.}~\bibnamefont
  {Held}},\ }\href {\doibase 10.1140/epjb/e2017-80115-7} {\bibfield  {journal}
  {\bibinfo  {journal} {Eur. Phys. J. B}\ }\textbf {\bibinfo {volume} {90}},\
  \bibinfo {pages} {114} (\bibinfo {year} {2017})}\BibitemShut {NoStop}%
\bibitem [{\citenamefont {Sekania}\ \emph {et~al.}(2017)\citenamefont
  {Sekania}, \citenamefont {Baeriswyl}, \citenamefont {Jibuti},\ and\
  \citenamefont {Japaridze}}]{Sekania:prb17}%
  \BibitemOpen
  \bibfield  {author} {\bibinfo {author} {\bibfnamefont {M.}~\bibnamefont
  {Sekania}}, \bibinfo {author} {\bibfnamefont {D.}~\bibnamefont {Baeriswyl}},
  \bibinfo {author} {\bibfnamefont {L.}~\bibnamefont {Jibuti}}, \ and\ \bibinfo
  {author} {\bibfnamefont {G.~I.}\ \bibnamefont {Japaridze}},\ }\href {\doibase
  10.1103/PhysRevB.96.035116} {\bibfield  {journal} {\bibinfo  {journal} {Phys.
  Rev. B}\ }\textbf {\bibinfo {volume} {96}},\ \bibinfo {pages} {035116}
  (\bibinfo {year} {2017})}\BibitemShut {NoStop}%
\bibitem [{\citenamefont {Du}\ \emph {et~al.}(2017{\natexlab{b}})\citenamefont
  {Du}, \citenamefont {Huang},\ and\ \citenamefont {Fiete}}]{Du:prb17c}%
  \BibitemOpen
  \bibfield  {author} {\bibinfo {author} {\bibfnamefont {L.}~\bibnamefont
  {Du}}, \bibinfo {author} {\bibfnamefont {L.}~\bibnamefont {Huang}}, \ and\
  \bibinfo {author} {\bibfnamefont {G.~A.}\ \bibnamefont {Fiete}},\ }\href
  {\doibase 10.1103/PhysRevB.96.165151} {\bibfield  {journal} {\bibinfo
  {journal} {Phys. Rev. B}\ }\textbf {\bibinfo {volume} {96}},\ \bibinfo
  {pages} {165151} (\bibinfo {year} {2017}{\natexlab{b}})}\BibitemShut
  {NoStop}%
\bibitem [{\citenamefont {Eckstein}\ and\ \citenamefont
  {Kollar}(2010)}]{Eckstein:njp10}%
  \BibitemOpen
  \bibfield  {author} {\bibinfo {author} {\bibfnamefont {M.}~\bibnamefont
  {Eckstein}}\ and\ \bibinfo {author} {\bibfnamefont {M.}~\bibnamefont
  {Kollar}},\ }\href {\doibase 10.1088/1367-2630/12/5/055012} {\bibfield
  {journal} {\bibinfo  {journal} {New J. Phys.}\ }\textbf {\bibinfo {volume}
  {12}},\ \bibinfo {pages} {055012} (\bibinfo {year} {2010})}\BibitemShut
  {NoStop}%
\end{thebibliography}%


%
\end{document}